# BiXiao: An AI-Based Atmospheric Environment Forecasting Model Using Discontinuous Grids

Shengxuan JI[1,5], Yawei QU[2*], Cheng YUAN[3*], Tijian WANG[1*], Bing LIU[4], Lili ZHU[4], Huihui ZHENG[4], Zhenfeng QIU[5], Pulong CHEN[6]

1. School of Atmospheric Science, Nanjing University, Nanjing 210023, China
2. College of Intelligent Science and Control Engineering, Jinling Institute of Technology, Nanjing, 211169, China
3. School of Emergency Management, Nanjing University of Information Science & Technology, Nanjing, 210044, China
4. China National Environmental Monitoring Center, Beijing 100012, China
5. FuYao Intelligence (Beijing) Technology Co., Ltd., Beijing 101108, China
6. Net Zero Era (Jiangsu) Environmental Technology Co., Ltd., Nanjing 210023, China
* Corresponding Author, E-mail: yawei.qu@jit.edu.cn
* Corresponding Author, E-mail: cyuan@nuist.edu.cn
* Corresponding Author, E-mail: tjwang@nju.edu.cn

**Abstract:** Currently, the technique of numerical model-based atmospheric environment forecasting has becoming matured, yet traditional numerical prediction methods struggle to balance computational costs and forecast accuracy, facing developmental bottlenecks. Recent advancements in artificial intelligence (AI) offer new solutions for weather prediction. However, most existing AI models does not have atmospheric environmental forecasting capabilities, while those with related functionalities remain constrained by grid-dependent data requirements, thus unable to deliver operationally feasible city-scale atmospheric environment forecasts. Here we introduce 'BiXiao', a novel discontinuous-grid AI model for atmospheric environment forecasting. 'BiXiao' couples meteorological and environmental sub-models to generate predictions using site-specific observational data, completing 72-hour forecasts for six major pollutants across all key cities in the Beijing-Tianjin-Hebei region within 30 seconds. In the comparative experiments, the 'BiXiao' model outperforms mainstream numerical models in both computational efficiency and forecast accuracy. It surpasses CAMS with respect of operational 72-hour forecasting and exceeds WRF-Chem's performance in heavy pollution case predictions. The 'BiXiao' shows potential for nationwide application, providing innovative technical support and new perspectives for China's atmospheric environment forecasting operations.

# 1 Introduction

Numerical models have long been an essential method for atmospheric environment research and have become one of the key technical approaches for atmospheric environmental protection in recent years in China. According to modeling theory, atmospheric environment models can be broadly classified into two categories: statistical models and numerical models. Statistical models are centered around statistical algorithms such as regression, classification, and fitting, and they predict the future evolution of atmospheric pollutant concentrations by analyzing patterns within the data. In contrast, numerical models are based on fundamental atmospheric environmental theories and use meteorological fields as drivers to simulate real atmospheric environmental changes through mathematical systems, ultimately predicting the spatiotemporal variations of atmospheric pollutant concentrations.

In the 1990s, with the rapid development of high-performance computing technology, numerical weather prediction (NWP), based on CPU computations, gradually became the mainstream approach. Atmospheric environment models entered the stage of complex process-based numerical models. These models further integrated meteorology, atmospheric physics, atmospheric chemistry, mathematics, and computer science, describing the dynamics and physical-chemical nonlinear processes in the atmosphere through numerical computations. Additionally, by incorporating numerous chemical equations, they accounted for the transformation between multiple species, thus simulating the atmospheric composition and pollution processes (Wang et al., 2008).

In recent years, researchers have increasingly recognized that combining atmospheric environment models with meteorological models—utilizing meteorological fields to compute parameters such as turbulence and boundary layers—can significantly enhance the model's ability to simulate pollutant dispersion. The development of atmospheric environment models has now entered the "meteorology-chemistry" coupling model phase. Coupling regional or global meteorological models with atmospheric chemical models has become the primary approach in atmospheric environmental numerical simulations. In this coupled framework, the meteorological model module mainly simulates atmospheric dynamics, thermodynamics, and other physical processes, while the atmospheric chemical model module focuses on simulating chemical reactions in the atmosphere, as well as the emission, transport, diffusion, and deposition of pollutants (Wang et al., 2024).

At the beginning of the 21st century, researchers in China began to independently develop or optimize atmospheric environment models. Institutions such as the Institute of Atmospheric Physics of the Chinese Academy of Sciences, the China Meteorological Administration, Nanjing University, and Southern University of Science and Technology have successively released a series of atmospheric environment forecasting models, including the Nested Air Quality Prediction Model System (NAQPMS), the Chemical Weather Forecast System (CUACE), the Regional Atmospheric Environmental Model System (RegAEMS), and the Regional Meteorology-Chemistry Coupling Model (WRF-GC) (Wang et al., 2006; Zhou et al., 2012; Wang et al., 2012; Lin et al., 2020). China has begun to establish and gradually improve its operational atmospheric environment forecasting system, with relevant model-based systems being promoted for use in cities and regions across the country.

It is noteworthy that the models used in atmospheric environment forecasting are all based on traditional numerical models. These models typically rely on a physical understanding of weather processes, translating complex atmospheric motions and physical processes into mathematical models. A computational framework is then built upon this foundation to solve the corresponding system of partial differential equations via high-performance computers, ultimately providing predictions of future weather and atmospheric environment states. Despite significant progress in forecast accuracy due to continuous advancements in high-performance computing technology and deeper research into atmospheric dynamics and physical processes, these numerical models still face substantial technical bottlenecks in terms of performance and computational cost. This is particularly evident when meeting future demands for shorter, higher-frequency forecasts, as their relatively complex computational processes continue to present challenges.

In recent years, with the rapid development of Artificial Intelligence (AI), new approaches for obtaining faster and more accurate weather forecasting results have emerged (Düben et al., 2021). Among these, AI-driven large meteorological models (hereafter referred to as "large models") have introduced a groundbreaking new paradigm in meteorological research, characterized by "data-driven" techniques. Unlike traditional Numerical Weather Prediction (NWP), which relies on physical mechanisms for modeling, large models use deep learning algorithms to learn complex nonlinear relationships from massive historical meteorological data, thereby constructing neural network models for weather prediction. After sufficient training, these large models can recognize key changing trends in the atmospheric system, explore the evolution of weather phenomena at the data level, and directly apply this knowledge for weather forecasting (Huang et al., 2024).

During the inference phase, large models typically require only a single Graphics Processing Unit (GPU) to complete predictions, significantly reducing computational costs compared to traditional models that depend on large-scale Central Processing Unit (CPU) clusters. At the same time, their computational efficiency is greatly improved, enabling a dramatic reduction in the time required to complete forecasts with the same spatial and temporal resolution. For example, the "Pangu" AI weather model developed by Huawei Cloud in China can produce global weather forecasts for the next 7-10 days within minutes (Bi et al., 2023).

In addition to forecasting speed, large models also have notable advantages in forecast spatial resolution. Google's "MetNet" series models have achieved spatial and temporal resolutions of 1 km and 30 seconds, respectively, surpassing the National Oceanic and Atmospheric Administration's (NOAA) High-Resolution Rapid Refresh (HRRR) system in precipitation forecasts over a 7-8 hours period (Andrychowicz et al., 2023; Sønderby et al., 2020). The development of large models provides faster decision support for extreme weather and disaster forecasting and emergency response. Furthermore, their higher computational efficiency makes large-scale ensemble forecasting feasible. For instance, Google's GenCast approach can optimize the predictive capability of large models through ensemble methods (Price et al., 2023).

In June 2024, the China Meteorological Administration launched the AI-based global short- to medium-term forecast system "Fengqing", the AI-based nowcasting system "Fenglei", and the AI-based global subseasonal-to-seasonal prediction system "Fengshun." This was followed by the initiation of the "Artificial Intelligence Weather Forecast Large Model Demonstration Plan (AMI-FDP)," marking the rapid development of meteorological large models in China (Huang et al., 2024; Huang et al., 2024).

The environmental large model is an extension of the meteorological large model and represents an important branch in the future development of meteorological large models in the field of atmospheric environment. However, most of the large models that have been released so far do not involve atmospheric chemistry, and therefore cannot yet forecast atmospheric environmental parameters. In June 2024, Microsoft launched the "Aurora" large model, which for the first time included simulations of atmospheric environment components. Based on the European Centre for Medium-Range Weather Forecasts (ECMWF) Copernicus Atmosphere Monitoring Service (CAMS) analysis fields, it achieved global atmospheric environment forecasting. "Aurora" is capable of simulating 5-day global atmospheric environment forecasts and 10-day global weather forecasts within 1 minute (Bodnar et al., 2024).

However, the "Aurora" model still has significant shortcomings in its forecast results. First, aside from $PM_{2.5}$ and $PM_{10}$, which has ground-level concentration forecasts, other pollutants only provide column concentration forecasts, greatly reducing the practical utility of the forecast results. Additionally, the spatial resolution of its atmospheric environment forecast is only 0.4°, which is insufficient for fine-scale simulation of urban environmental parameters. In fact, it is clear that the deficiencies in "Aurora's" atmospheric environment forecasting stem from its heavy reliance on reanalysis data. The types of pollutants that can be forecasted and the resolution of these forecasts must be consistent with the CAMS analysis data, thus limiting its practical application in air quality forecasting. This issue is a common limitation of current large models. Existing large models are primarily trained based on gridded reanalysis data (e.g., ERA5), which leads to two problems: firstly, the elements that the model can simulate are often limited to those provided by the reanalysis data; and secondly, the model's forecasting ability is relatively low in cases where training data for high-impact weather or severe pollution events is insufficient.

Although some large models have attempted improvements, such as the Fuxi model, which has recently used satellite data to drive its forecasts, most meteorological large models still cannot directly integrate multi-source meteorological observation data, including ground-based observations, upper-air observations, and radar data, for forecasting purposes (Sun et al., 2024).

Since 2013, China has made significant breakthroughs in atmospheric environmental research. With the rapid pace of urbanization, air pollution control strategies have gradually shifted from "weather-driven" to "emission control." The increasingly stringent atmospheric environmental regulations have raised higher demands for future atmospheric environmental forecasting (Meng et al., 2023, 2025). However, current numerical models struggle to meet the forecasting demands of China's rapid development, and the large models that have been released so far still fail to provide atmospheric environment forecast products suitable for operational use. Therefore, there is an urgent need to develop a large model capable of atmospheric environment forecasting, which can utilize China's extensive atmospheric environment monitoring network data to provide accurate, fast, and cost-effective forecast products. This would support the national and regional demands for "refined, precise, and accurate" atmospheric environment forecasting.

To address this need, this paper proposes a novel AI-driven atmospheric environment forecasting model, "BiXiao" (hereafter referred to as "BiXiao"). Drawing on the successful experience of traditional atmospheric environment numerical modeling systems, "BiXiao" innovatively introduces an "discontinuous grid" design, enabling it to effectively integrate surface observational data for particulate matter, gaseous pollutants, and other atmospheric contaminants. This allows for a comprehensive forecast capability across all atmospheric environment parameters. "BiXiao" can accurately forecast atmospheric conditions for different cities and regions, offering efficient computation, precise forecasts, and an exploration of underlying mechanisms, while also considering operational efficiency and cost. It can be integrated with traditional methods, providing a new technological pathway for atmospheric environment forecasting.

## 2 Model Introduction

### 2.1 Name of Model

"BiXiao" is an elegant ancient term for "blue sky," originating from the famous verse in Liu Yuxi's poem. The model is named "BiXiao" to symbolize the aspiration for a beautiful atmospheric environment. It is hoped that this atmospheric environment large model will support the nation's "pollution reduction and carbon reduction" strategy, continuously evolve and innovate, achieve technological breakthroughs, and provide strong support for air pollution prevention and control.

### 2.2 Model Architecture

#### 2.2.1 General Architecture

The "BiXiao" model adopts an "offline" architecture, consisting of two non-coupled modules: the Meteorological Module and the Environmental Module. The combination of these two modules is similar to the Stand-Alone mode of the WRF-CMAQ atmospheric environment numerical forecast model, where the output of the environmental module does not feedback into the meteorological module's computations. This design ensures independence and modularity, allowing for flexible integration and potential updates to either module without affecting the overall system.

In the specific structural design of the modules, "BiXiao" addresses the limitations of traditional AI large models, which often excessively depend on regularly gridded data. To better adapt to the current state of atmospheric environmental observation data, "BiXiao" employs a heterogeneous model structure. Specifically, the meteorological module retains the traditional gridded structure used in Numerical Weather Prediction (NWP) models, which is well-suited for simulating large-scale atmospheric dynamics. In contrast, the environmental module adopts a discrete grid structure. This approach does not rely on complete, regularly gridded data; instead, it can effectively use scattered observational data from various monitoring stations. This flexibility enables the environmental module to be trained and run using localized, real-time environmental data, making the model more adaptable to the actual data availability and improving its capability to simulate regional and urban air quality dynamics.

During the training phase, the "BiXiao" model's Meteorological and Environmental Modules are trained separately. The meteorological large model is an autoregressive model, which can be trained using reanalysis data such as ERA5. The environmental module is trained under ideal meteorological conditions, using ERA5 or similar reanalysis data as the meteorological background, along with atmospheric environment station observation data for model training.

In the inference phase, the model's forecasting task is completed through sequential collaboration between the two modules. First, the meteorological module calculates the gridded meteorological forecast field, which is used to drive the meteorological background for the environmental module. Then, the environmental module combines the meteorological field data with discrete environmental observation data to output pollutant concentration forecasts on the discrete grid. Finally, the initial conditions from the meteorological module at time T+0, along with the T+1 forecast obtained through a one-step inference, are used as input for the environmental module's meteorological input. This is combined with the discrete grid environmental data at T+0, which is then input into the environmental module to achieve the T+1 pollutant concentration prediction on the discrete grid.

**2.2.2 Meteorological Module**

The meteorological module of "BiXiao" adopts the 3D Swin Transformer (hereafter referred to as Swin3D) as its backbone network. The Swin Transformer is a vision model based on the Transformer architecture, where the core idea is to divide the input image into multiple windows and apply the self-attention mechanism within each window to process the image (Liu et al., 2021). Compared to the traditional 2D Swin Transformer (Swin2D), Swin3D is better equipped to handle an additional dimension. While Swin2D focuses on extracting data features within a two-dimensional plane, Swin3D extends to three-dimensional space, enabling it to handle data with highly variable features, which aligns with the three-dimensional structure of the atmosphere. Thus, Swin3D is well-suited for building atmospheric environment models, and this approach has been effectively applied in Huawei's Pangu-Weather large meteorological model (Bi et al., 2023).

The training objective of the meteorological module is to predict the meteorological field at time T+1 based on the meteorological element data at time T+0, as illustrated in **Figure 1a**. In the specific model structure, to leverage Swin3D for extracting the spatiotemporal features of meteorological data, the surface field and pressure level field data at T+0 are first processed through patch embedding. This involves applying 2D convolution (Conv2D) to the surface field features, while the pressure level field data undergoes 3D convolution (Conv3D). Subsequently, the convolutional feature data of both the surface field and pressure level fields are concatenated, forming a three-dimensional feature vector.

The Swin3D model uses a "U-shaped" architecture to extract key data features. This is done by first downsampling to reduce information density, followed by Swin3D for key feature extraction, and then upsampling to increase information density. Finally, a patch recovery process is applied, using deconvolution to restore the feature data, generating the ground and height field meteorological data for time T+1.

In essence, the training of the meteorological module involves using neural network to learn the mapping from the three-dimensional atmospheric at T+0 to the corresponding features at T+1. The model parameters are optimized through the use of numerous paired training samples of T+0 and T+1, thereby improving forecast accuracy.

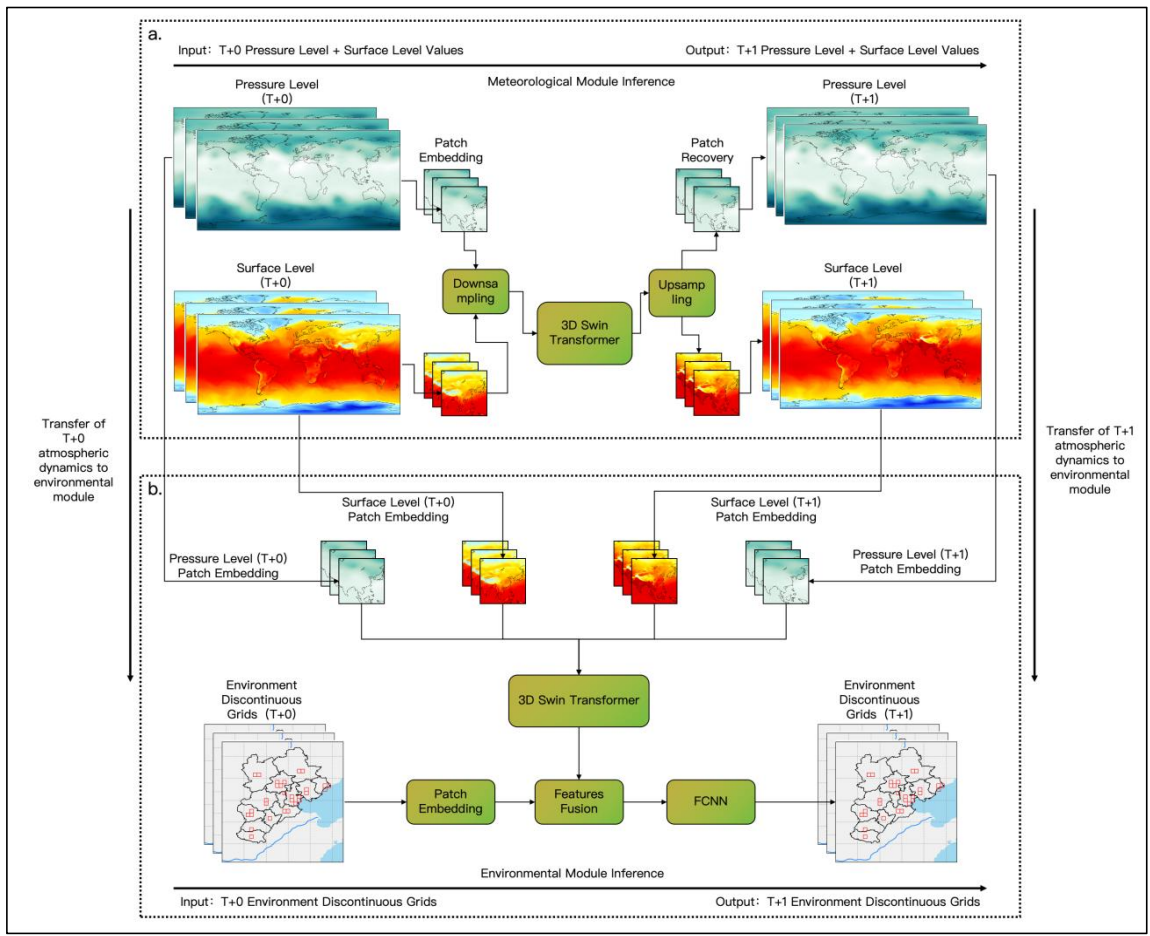

**Figure 1:** Schematic Diagram of the "BiXiao" Large Model Architecture

### 2.2.3 Environmental Module

Compared to the meteorological module, the structure of the environmental module is more complex. Unlike the meteorological module, which predicts the meteorological field at T+1 based solely on T+0 meteorological data, the environmental module must simultaneously use both T+0 meteorological and environmental field data, along with T+1 meteorological field data, to infer the environmental field at T+1. It is important to note that environmental field data is not regular gridded data but instead discrete grid data derived from station-based observational data.

Within the environmental module, Swin3D is first employed as the backbone network to extract the three-dimensional meteorological field features at both T+0 and T+1, providing the atmospheric dynamic conditions for predicting environmental elements (as shown in **Figure 1b**). This process is overall similar to the feature extraction of raw meteorological field data in the meteorological module, with two key differences:

**Dual Time-Step Processing:** While the meteorological module only processes the meteorological elements at T+0, the environmental module needs to process both T+0 and T+1 meteorological field data simultaneously.

**Feature Retention and Fusion:** In the meteorological module, the features extracted by Swin3D are subjected to upsampling and patch recovery to generate a complete meteorological field. In contrast, the environmental module does not perform downsampling, and the atmospheric dynamic feature information processed by Swin3D is directly retained for subsequent fusion with environmental information.

Subsequently, the discrete grid environmental data at T+0 is processed through an encoding network, and the extracted meteorological features are mixed with this encoded environmental data. A fully connected neural network then completes the prediction of the discrete grid environmental data at T+1.

From an overall process perspective, the core idea of the environmental module is as follows: Given the known conditions of the meteorological fields at T+0 and T+1, as well as the environmental field at T+0, the module first extracts the dynamic conditions and evolution characteristics of the meteorological fields at both T+0 and T+1. These are used to construct the atmospheric dynamic "background" or "forcing" information required for environmental element prediction. This atmospheric feature information is then fused with the environmental information at T+0 to predict the environmental information at T+1.

## 3 Model Training and Validation

### 3.1 Data and Study Area

#### 3.1.1 Study Area

In this study, the Beijing-Tianjin-Hebei (hereafter referred to as BTH) region (34.75 °N –43 °N, 112.5 °E –120.75 °E) is selected as the research area. This region is one of the most economically developed and densely populated in China. It is also the core battleground of China's "Blue Sky Protection Campaign" and a key area for atmospheric environmental protection. Furthermore, it is one of the first regions in China to deploy atmospheric environment forecasting services. In the future, precise forecasting of air pollutants in the BTH region will continue to be of significant importance, and atmospheric environment forecasting remains a foundational task for environmental protection in the area.

The BTH region is a high-incidence area for haze events, and its pollution characteristics are representative of other major urban areas in China. Conducting model training and testing in this region will help improve the model's adaptability and provide a reliable basis for its broader application across the country. Moreover, compared to direct nationwide training and forecasting, focusing on a regional study reduces the computational resource demands, thus facilitating the efficient development and optimization of the model.

**3.1.2 Meteorological Field Data**

The "BiXiao" model uses ERA5 reanalysis data as the driving data for the meteorological module training. ERA5 is generated using the four-dimensional variational (4D-Var) data assimilation and model forecasting methods from the ECMWF Integrated Forecasting System (IFS), version CY41R2. The data has a horizontal resolution of 0.25°×0.25° and is widely used in atmospheric science (Hersbach et al., 2020). The ERA5 data used in this study includes both three-dimensional and two-dimensional variables.

- Three-Dimensional Variables: These primarily include vertical velocity, temperature, and geopotential height, with values obtained from a global model on 137 model layers. The data is then interpolated through the FULL-POS method in IFS to 37 pressure levels.

- Two-Dimensional Variables: These include surface (or single-layer) data, such as precipitation and top-of-atmosphere radiation, with values provided as vertical integrals over the entire atmospheric depth.

In terms of temporal coverage, this study uses ERA5 data from January 2014 to March 2024, extracting meteorological variables at UTC times 00, 06, 12, and 18. Since the primary task of the environmental module is to predict various chemical environmental elements at ground-based observation stations, this study focuses on six high-altitude atmospheric variables at four pressure levels (500, 850, 925, and 1000 hPa), as well as five surface variables, totaling 29 elements. These specific variables are listed in **Table 1**.

Table 1: List of Meteorological Data for the BiXiao Meteorological Module

| short-name | long-name | levels |
|---|---|---|
| **u** | U-component of wind | 500、850、925、1000 hPa |
| **v** | V-component of wind | 500、850、925、1000 hPa |
| **w** | Vertical velocity | 500、850、925、1000 hPa |

| t | Temperature | 500、850、925、1000 hPa |
|---|---|---|
| q | Specific humidity | 500、850、925、1000 hPa |
| z | Geopotential | 500、850、925、1000 hPa |
| t2m | 2m temperature | surface |
| u10m | 10m u-component of wind | surface |
| v10m | 10m v-component of wind | surface |
| d2m | 2m dewpoint temperature | surface |
| sp | Surface pressure | surface |

In terms of data area selection, considering that this study focuses on the BTH region and to improve training efficiency, the original ERA5 data with a spatial resolution of 0.25° (a total of 721×1440 grid points) is cropped. The data driving the meteorological module covers the area from 14.75°N to 55°N and 65°E to 140°E, as shown in the full map in **Figure 2 (left)**, corresponding to 160×300 grid points. The data driving the environmental module is further focused on the BTH region (34.75°N–43°N, 112.5°E–120.75°E), corresponding to 34×34 grid points, as shown in **Figure 2 (right)**.

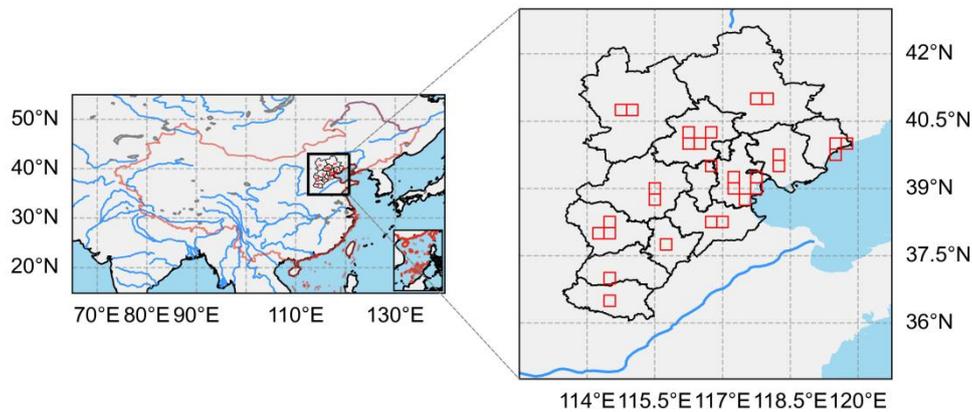

**Figure 2:** Meteorological Data Range for the Meteorological Module (Left) and Meteorological Data Range and Environmental Grid Distribution for the Environmental Module (Right)

### 3.1.3 Environmental Field Data

This study uses hourly observational data from 79 environmental monitoring stations in the BTH region, covering the period from January 2021 to March 2024, as training data. The observational data includes the six major pollutants specified in China's "Ambient Air Quality Standards," namely sulfur dioxide ($SO_2$), nitrogen dioxide ($NO_2$), carbon monoxide (CO), ozone ($O_3$), and particulate matter ($PM_{2.5}$ and $PM_{10}$).

To ensure compatibility with the ERA5 meteorological data at a 0.25 ° resolution, the station observation data is gridded. Specifically, each station's observational data is assigned to the nearest ERA5 grid based on the station's latitude and longitude. For grids with only one station, the atmospheric environmental observation value directly uses the data from that station. If multiple monitoring stations exist within a grid, the air quality level of the grid is represented by the average value of all station data. After data processing, the 79 stations in the BTH region are mapped to 29 valid grids, as shown by the red grid points in **Figure 2 (right)**.

### 3.2 Model Training

Before training the model, all input data, including ERA5 meteorological fields and environmental station data, need to undergo preprocessing. The preprocessing steps mainly include calculating the mean and variance required for data normalization, ensuring the stability of data distribution during the model training process. After preprocessing, all data are divided into two parts: the "training set" for training and the "test set" for testing.

For the selection of training data, the meteorological module's training data include ERA5 data from January 2014 to March 2024, using data from four time points: UTC 00, 06, 12, and 18. The environmental module's training data use hourly data from January 2021 to March 2024. For the test set selection, to ensure objectivity, five consecutive days of data are randomly selected each month as the test set, and this portion of data does not participate in model training and validation. Additionally, to further evaluate the model's performance in forecasting typical heavy pollution events, extra data from an $O_3$ pollution event in the BTH region in June 2022 and a $PM_{2.5}$ pollution event in October 2023 are included in the "test set," while the remaining data are used as the training set.

The model training uses smoothL1 loss as the loss function and employs supervised learning to optimize the model's single-step autoregressive prediction. The model construction and training are based on the PyTorch framework, and the AdamW optimizer is used in conjunction with a linear decay scheduler to dynamically adjust the learning rate, improving training convergence efficiency. The meteorological module and environmental module are trained independently, with substantial differences in training duration and computational resource requirements. For example, the meteorological module is trained using 8 Nvidia RTX 4090 GPUs, completing 150 epochs of training in several hours, with the final loss reduced to 0.02. The environmental module is trained on a single Nvidia RTX 4090 GPU, requiring 24 hours to complete 200 epochs of training, with the final loss reduced to 0.01. While further training could still lower the loss, the prediction accuracy on the validation set did not show significant improvement, indicating that a lower loss does not necessarily correlate with better generalization ability.

### 3.3 Model Validation

#### 3.3.1 Validation Metrics

The model performance is evaluated using three metrics: Root Mean Square Error (RMSE), Mean Absolute Error (MAE), and Pearson Correlation Coefficient (PCC).

- RMSE reflects the average deviation between the model's predicted values and the actual values. It is more sensitive to larger errors and provides a comprehensive measure of the overall deviation of the predicted values.
- MAE is used to assess the average absolute error between the predicted and actual values, focusing on the actual magnitude of errors.
- PCC measures the linear correlation between the predicted values and the actual values. A PCC value closer to 1 or -1 indicates a strong linear correlation between the predicted and actual values, while a value closer to 0 indicates a weak linear correlation.

These three metrics are common model evaluation parameters and have been widely used in many related studies. The specific calculation methods for the three parameters are as follows:

$$RMSE(c,t) = \sqrt{\frac{1}{n}\sum_{i=1}^{n}(y_{c,t,i} - \hat{y}_{c,t,i})^2} \quad (1)$$

$$MAE(c,t) = \frac{1}{n}\sum_{i=1}^{n}|y_{c,t,i} - \hat{y_{c,t,i}}| \quad (2)$$

$$PCC(c,t) = \frac{\sum_{i=1}^{n}(y_{c,t,i} - \bar{y_{c,t,i}})(\hat{y_{c,t,i}} - \bar{y_{c,t,i}})}{\sqrt{\sum_{i=1}^{n}(y_{c,t,i} - \bar{y_{c,t,i}})^2 \sum_{i=1}^{n}(\hat{y_{c,t,i}} - \bar{y_{c,t,i}})^2}} \quad (3)$$

Where c represents the evaluated element; t represents the forecast time; and n is the number of samples.

### 3.3.2 Validation Experiment

As mentioned earlier, to avoid having data samples in the test set that are similar to those in the training set, this study strictly divides the training and test datasets to ensure their independence. In the validation experiment, simulation predictions are made for the selected data samples from the test set, and the results are compared with the measured data. For the 5-day test set, ERA5 reanalysis data for the first two days at UTC 00, 06, 12, and 18 hours are used as the initial input for the meteorological fields (T+0), and the corresponding environmental observation data for the same times are used as the initial condition of the environmental fields (T+0). The model uses a 6-hour time step for inference, generating pollution concentration forecasting results for the next 72 hours. The forecasting results are then compared with the remaining 3 days of measured data. The entire validation experiment includes 308 forecast times, and the model's prediction results are verified using the measured data. The RMSE, MAE, and PCC values for the forecast results at different time steps on all grids are calculated and used to assess the model's forecasting ability.

**Figure 3** shows the overall forecast performance of the "BiXiao" model. The forecast results are categorized and statistically analyzed according to the latitude and longitude of the grids and the administrative divisions of the BTH region. The analysis focuses on the forecast correlation coefficients (PCC) for $O_3$, $PM_{2.5}$, and $PM_{10}$ at 6 hours, 48 hours, and 72 hours forecast times for different administrative regions. **Figure 4** presents the PCC, RMSE, and MAE for six pollutants ($O_3$, $PM_{2.5}$, $PM_{10}$, $NO_2$, CO, $SO_2$) at different forecast time steps across all grids.

Among the six pollutants, "BiXiao" performs best in $O_3$ forecasts. For forecasts from 6 h to 72 h, the PCC for $O_3$ consistently outperforms other pollutants, with the average $O_3$ PCC across 29 grids reaching as high as 0.91 at 6 h. The RMSE and MAE for $O_3$ forecasts remain moderate, at 26±6 μg/m³ and 20±5 μg/m³, respectively. Regional analysis shows that the forecast performance for $O_3$ is better in the central areas of BTH, such as Beijing, Tianjin, Langfang, and Baoding, where the PCC at all forecast times is slightly higher than in other areas of the region.

For particulate matter concentration forecasts, the performance of the "BiXiao" model for $PM_{2.5}$ and $PM_{10}$ is similar, with slightly better performance for $PM_{2.5}$. At the 6-hour forecast, the PCC for $PM_{2.5}$ and $PM_{10}$ are 0.86 and 0.79, respectively. As the forecast time increases, the correlation gradually decreases, with the PCC for $PM_{10}$ showing a more pronounced drop after 48 h and remaining slightly lower than for $PM_{2.5}$. Since $PM_{10}$ concentration includes $PM_{2.5}$, the RMSE and MAE for $PM_{10}$ are higher than those for $PM_{2.5}$ across all forecast periods, with a noticeable increase in RMSE and MAE for $PM_{10}$ after 48 hours. The reasons behind this trend require further investigation. Spatial distribution analysis indicates that "BiXiao" performs better in simulating particulate matter in the southern and central regions of BTH.

For other gaseous pollutants, the "BiXiao" model shows acceptable forecast ability for $NO_2$ and CO, with a relatively steady decrease in PCC as the forecast time increases. The RMSE and MAE for these two pollutants also increase gradually with valid time, though the increase is relatively small. Specifically, the RMSE for $NO_2$ and CO are 14±3 μg/m³ and 0.32±0.04 mg/m³, respectively, while the MAE for $NO_2$ and CO are 11±2 μg/m³ and 0.22±0.04 mg/m³, respectively. Compared to other pollutants, the forecast correlation for $SO_2$ is the lowest, but its RMSE and MAE show relatively stable changes.

Overall, these results suggest that "BiXiao" excels in simulating $O_3$ and particulate matter, while there is still room for improvement in simulating $SO_2$.

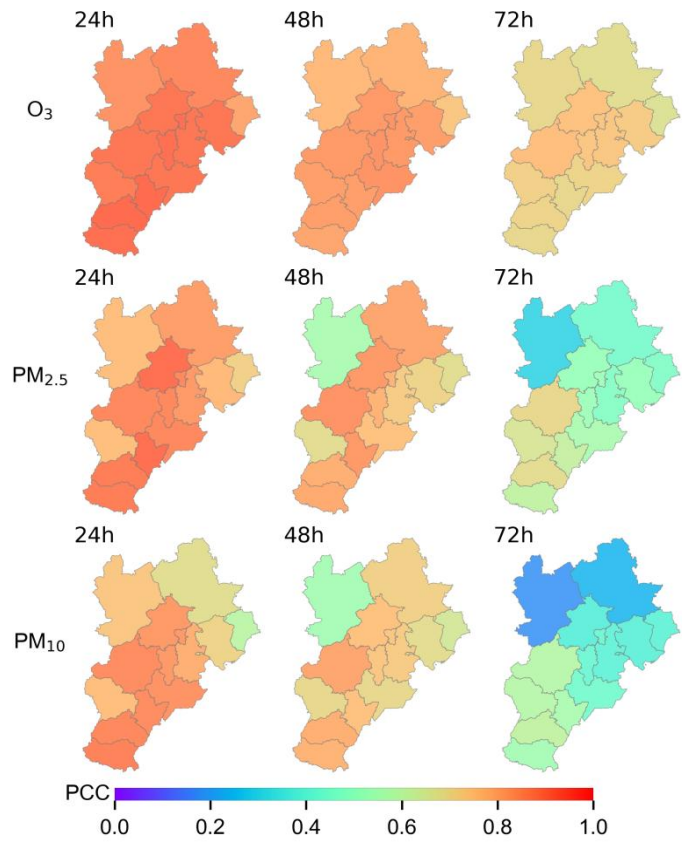

**Figure 3:** Pearson Correlation Coefficient (PCC) between the simulated and observed values for "BiXiao," where the first column represents the 24-hour forecast, the second column represents the 48-hour forecast, and the third column represents the 72-hour forecast.

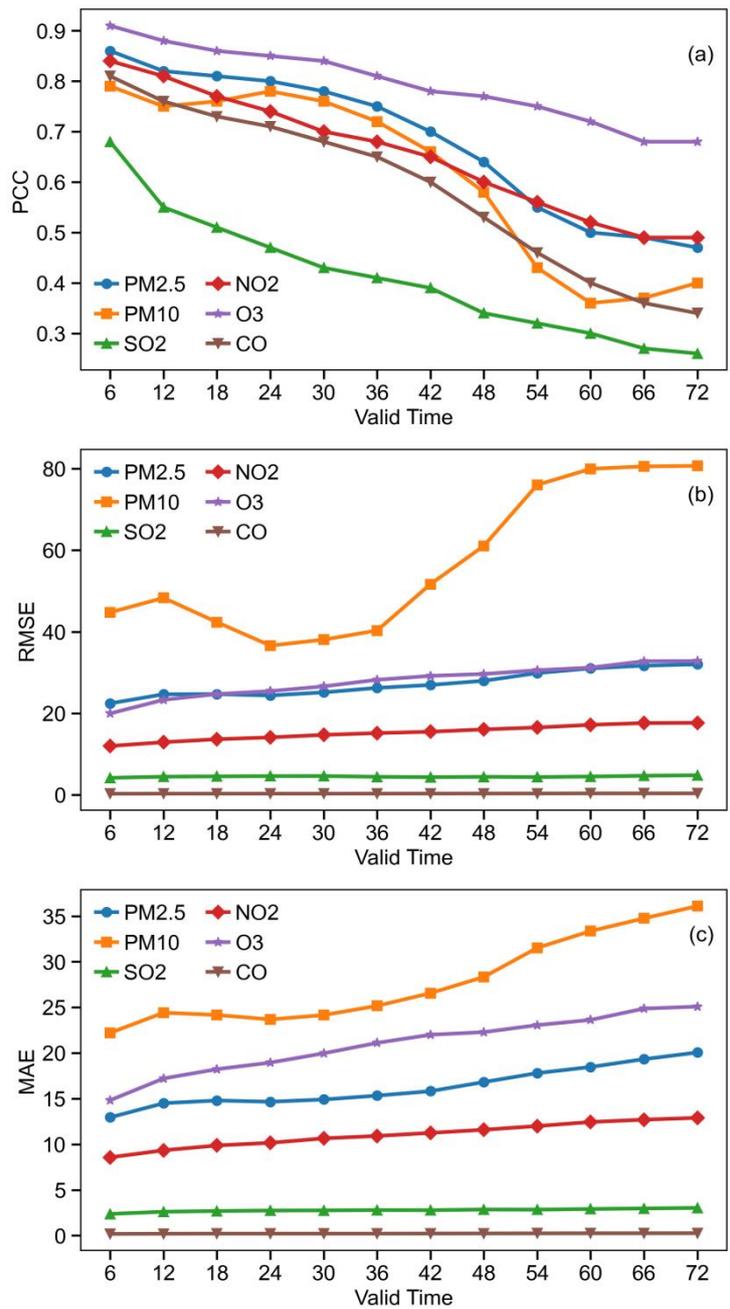

**Figure 4:** the following between the predicted values of "BiXiao" and the actual observed values:

(a) Pearson Correlation Coefficient (PCC)

(b) Root Mean Square Error (RMSE)

(c) Mean Absolute Error (MAE).

## 4 Routine Environmental Forecasting: "BiXiao" vs CAMS

## 4.1 Experimental Setup

ECMWF's CAMS provides global atmospheric environmental monitoring and forecasting, widely recognized for its high forecasting capability and broad representativeness (Inness et al., 2019). Therefore, this study selects CAMS forecast data as a benchmark to evaluate the performance of the "BiXiao" model in operational environmental forecasting.

When comparing, two main issues must be considered: spatial and temporal matching. In terms of spatial matching, environmental forecast results from CAMS are extracted from the corresponding locations based on the latitude and longitude of the 29 discrete environmental grids in "BiXiao." For temporal matching, both CAMS and BiXiao provide 6-hour forecast intervals, and the comparison is made over the 72-hour forecast period.

It is worth noting that since CAMS only provides forecast results at UTC 00:00 and 12:00 each day, the comparison is conducted using only the common start times between CAMS and BiXiao. Additionally, as CAMS provides surface concentrations of $PM_{2.5}$ and $PM_{10}$ only, the comparison is focused only on particulate matter ($PM_{2.5}$ and $PM_{10}$). The experiment uses observed environmental grid data to validate the forecast results from both CAMS and BiXiao to assess the differences in forecasting performance between the two models.

## 4.2 Experimental Results

### 4.2.1 Grid-based Statistical Comparison

The comparison of $PM_{2.5}$ and $PM_{10}$ forecast results between "BiXiao" and CAMS across the 29 valid grids is shown in **Figure 5**. In the 6-hour forecast, "BiXiao" outperforms CAMS, with a Pearson Correlation Coefficient (PCC) of 0.87 for $PM_{2.5}$ and 0.82 for $PM_{10}$, significantly higher than CAMS's PCC of 0.60 for $PM_{2.5}$ and 0.45 for $PM_{10}$. The RMSE (MAE) for $PM_{2.5}$ and $PM_{10}$ in "BiXiao" are 21.41 μg/m³ (12.45 μg/m³) and 41.55 μg/m³ (20.94 μg/m³), respectively, much lower than the RMSE (MAE) of 40.86 μg/m³ (27.7 μg/m³) and 72.03 μg/m³ (43.25 μg/m³) in CAMS.

As the valid time increases, the forecast performance of both models decreases. Overall, for most valid times, the PCC between the forecasted pollutant concentrations and observations decreases as forecast duration increases, with "BiXiao" maintaining higher correlation than CAMS. In the longer forecast period, i.e., the 72-hour forecast, "BiXiao" achieves PCC values of 0.44 for $PM_{2.5}$ and 0.40 for $PM_{10}$, while CAMS's PCC for $PM_{2.5}$ and $PM_{10}$ are 0.45 and 0.31, respectively.

From the error perspective, the RMSE and MAE for $PM_{2.5}$, as well as the MAE for $PM_{10}$, consistently show that "BiXiao" has lower deviations compared to CAMS across all forecast times. This indicates that "BiXiao" has an overall superior forecasting ability for particulate pollutants, particularly in the short-term forecasts.

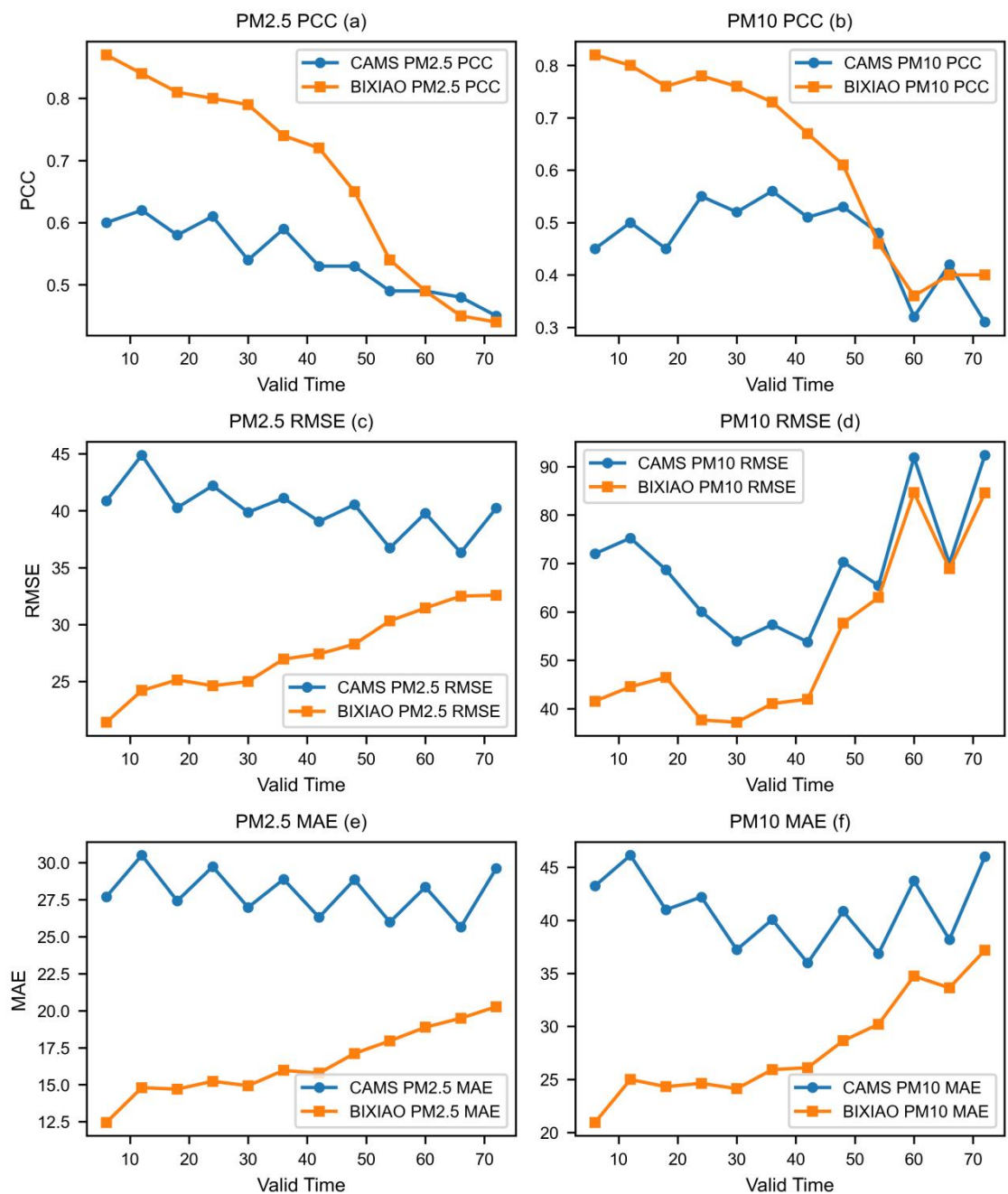

**Figure 5:** Comparison of forecast accuracy between the "BiXiao" model and CAMS:

(a) $PM_{2.5}$ Pearson Correlation Coefficient (PCC)

(b) $PM_{10}$ Pearson Correlation Coefficient (PCC)

(c) $PM_{2.5}$ Root Mean Square Error (RMSE)

(d) $PM_{10}$ Root Mean Square Error (RMSE)

(e) $PM_{2.5}$ Mean Absolute Error (MAE)

(f) PM$_{10}$ Mean Absolute Error (MAE)

**4.2.2 Effective Grid Comparison**

On the 29 effective grids, the forecast performance of the "BiXiao" model for PM$_{2.5}$ and PM$_{10}$ was compared with the numerical model CAMS at the same forecast time, as shown in **Figure 6**. The comparison results are presented using difference value, where the difference in the Pearson Correlation Coefficient (PCC) is calculated as "BiXiao" minus "CAMS," and the difference in the Root Mean Square Error (RMSE) is calculated as "CAMS" minus "BiXiao."

For PM$_{2.5}$ forecasting, within the first 48 hours, BiXiao consistently outperforms CAMS with higher PCC and smaller RMSE across all grids. After 48 hours, BiXiao's PCC is slightly lower than that of CAMS in the central and eastern regions of the BTH area, but the RMSE is still smaller than CAMS in most grids.

For PM$_{10}$ forecasting, the evaluation results in the first 24 hours show similar characteristics to those of PM$_{2.5}$, with BiXiao achieving higher PCC and smaller RMSE in all grids compared to CAMS. After 24 hours, BiXiao continues to outperform CAMS in PCC, and most regions also show smaller RMSE than CAMS, with a few grids in the southern region showing slightly worse performance. After 48 hours, BiXiao's PCC in the southern and northern regions is slightly lower than CAMS, with similar trends observed in the RMSE. This trend is speculated to be related to the method of combining environmental grids, where each environmental grid represents the average data of all the monitoring stations within that grid. In the central urban areas, where monitoring stations are more concentrated, the data in each grid is more stable, leading to better model performance in these areas.

Overall, BiXiao shows a significant advantage in the first 48 hours of forecasting. After 48 hours, the forecast ability in some regions gradually declines to the point where it becomes comparable to or slightly worse than CAMS.

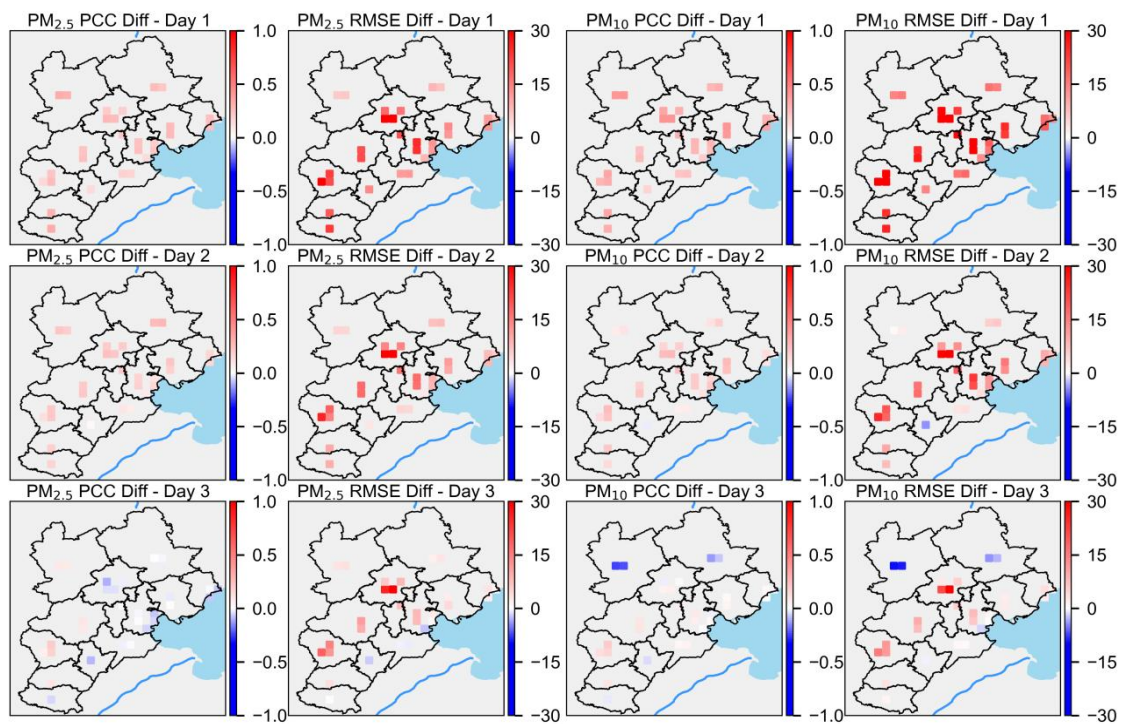

**Figure 6:** Comparison of daily average forecast performance for $PM_{2.5}$ and $PM_{10}$ between "BiXiao" and CAMS across different grids for 1-day, 2-day, and 3-day forecast periods: Pearson Correlation Coefficient (PCC) difference (BiXiao minus CAMS) and Root Mean Square Error (RMSE) comparison (CAMS minus BiXiao).

## 5 Heavy Pollution Case Forecast: "BiXiao" vs WRF-Chem

### 5.1 Experimental Setup

In this study, the WRF-Chem model, developed jointly by the NOAA Forecast Systems Laboratory (FSL) and the Pacific Northwest National Laboratory (PNNL), is selected as a representative atmospheric environment model for comparison. The analysis aims to compare the performance of WRF-Chem and the "BiXiao" model in predicting typical fine particulate matter ($PM_{2.5}$) and ozone ($O_3$) pollution events.

The two pollution events selected for this study are the $O_3$ pollution event from June 24–27, 2022, and the $PM_{2.5}$ pollution event from October 28–31, 2023. Both WRF-Chem and "BiXiao" models are used to simulate these two events, and the results are compared with observed data.

The WRF-Chem model is set up with two nested grids. The first grid consists of 91×74 horizontal grid points, while the second grid, which covers the BTH region, has a resolution of 27 km and includes 97×106 grid points. The boundary conditions for the second grid are provided by the first grid. The simulations for the two pollution events start at UTC 00:00 on June 21, 2022, and October 25, 2023, respectively. The first 72 hours are used for model integration, and the subsequent 72 hours serve as the effective forecast period.

In this study, WRF-Chem version 4.1 is used with the RADM2 chemical mechanism and the MADE/SOGARM aerosol parameterization scheme. Anthropogenic emission sources are based on the 2023 release of the Chinese Multiscale Emission Inventory (MEIC v1.4) (http://www.meicmodel.org/).

The "BiXiao" model uses ERA5 data for meteorological initial conditions at UTC 00:00 on June 24, 2022, and October 28, 2023, respectively, and uses environmental observational data from the same times as environmental initial conditions. The model runs with a 6-hour time step to forecast the next 72 hours. The experimental results are compared with observed environmental grid data to evaluate the forecast performance of both models during these typical pollution events.

### 5.2 Experimental Results

#### 5.2.1 Ozone Pollution Case Study

In June 2022, frequent extreme heat events occurred across northern China, coinciding with increased ozone ($O_3$) pollution. The temporal and spatial distribution of ozone pollution, including its intensity, significantly intensified during this period, with a strong correlation between temperature anomalies and $O_3$ pollution in the BTH region (Yang et al., 2025). This study focuses on a specific ozone pollution event from June 24 to 27, 2022, to assess the forecasting capability of the BiXiao model. During this event, peak $O_3$ concentrations in the BTH region occurred at 14:00 local time (UTC 06:00) on June 25 and 26, with grid-averaged $O_3$ concentrations reaching 270.0 μg/m³ and 163 μg/m³, respectively. During this period, the region was under the influence of a weather pattern following an upper-level trough (**Figure 7**), with clear skies, no precipitation, high temperatures, and weak winds. Afternoon surface temperatures in the grid averaged above 30 °C, peaking at 37°C on June 25, and the relative humidity at the surface was 76% in the afternoon of the same day. Strong sunlight and high temperatures accelerated photochemical reactions, fostering $O_3$ formation. The near-surface wind speed peaked at 4.8 m/s at 06:00 UTC on June 25 and 26, with wind speeds not exceeding 3.7 m/s at other times during the event, which hindered pollutant dispersion.

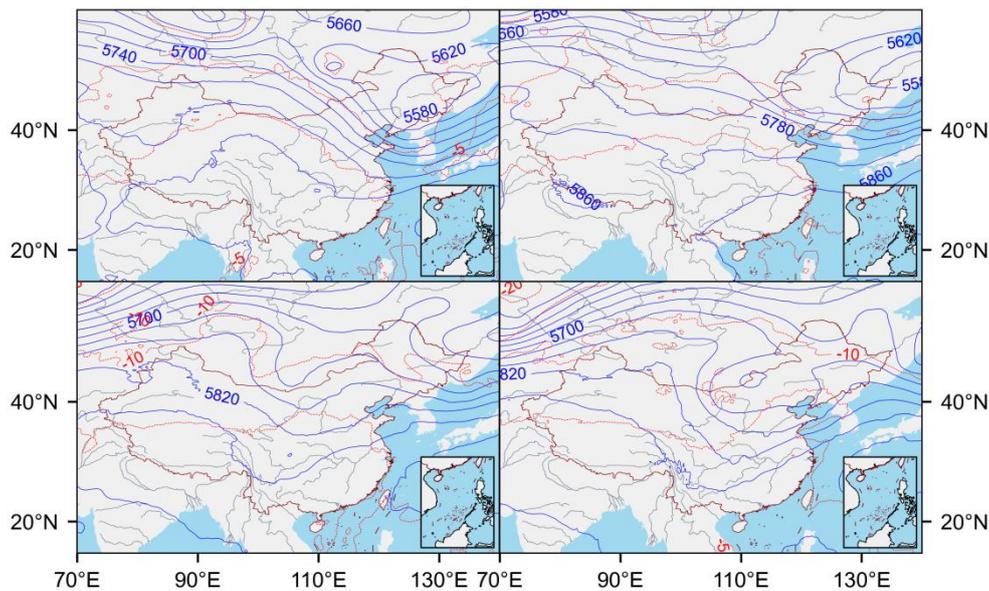

**Figure 7:** 500 hPa geopotential height and temperature map for June 24-27, 2022, 00:00 UTC.

**Figure 8** presents the forecast results of the BiXiao and WRF-Chem models, initialized at 00:00 UTC on June 24, 2022. Both models predict the trend of $O_3$ concentration well, though WRF-Chem tends to underestimate the values compared to observations. Throughout all forecast hours, the mean forecast values from BiXiao are closer to the observed values than those from WRF-Chem. During this pollution event, at 14:00 local time on June 25, 2022 (UTC 06:00), which corresponds to the 30th hour of the model forecast, $O_3$ concentration reached its peak, with a correlation coefficient of 0.82 between BiXiao and observations, while WRF-Chem had a value of -0.34. Six hours later, at the 36th hour of the forecast, $O_3$ concentration reached its secondary peak of 201 μg/m³, with a correlation coefficient of 0.55 for BiXiao and 0.02 for WRF-Chem. At these two times, the mean absolute errors of BiXiao were 85.33 μg/m³ and 56.37 μg/m³, respectively, lower than WRF-Chem's 181.38 μg/m³ and 91.28 μg/m³. For all other forecast times between 0 and 72 hours, the mean absolute error between BiXiao and observations was approximately 21.43 μg/m³, indicating the high accuracy of the BiXiao model in forecasting the ozone pollution event.

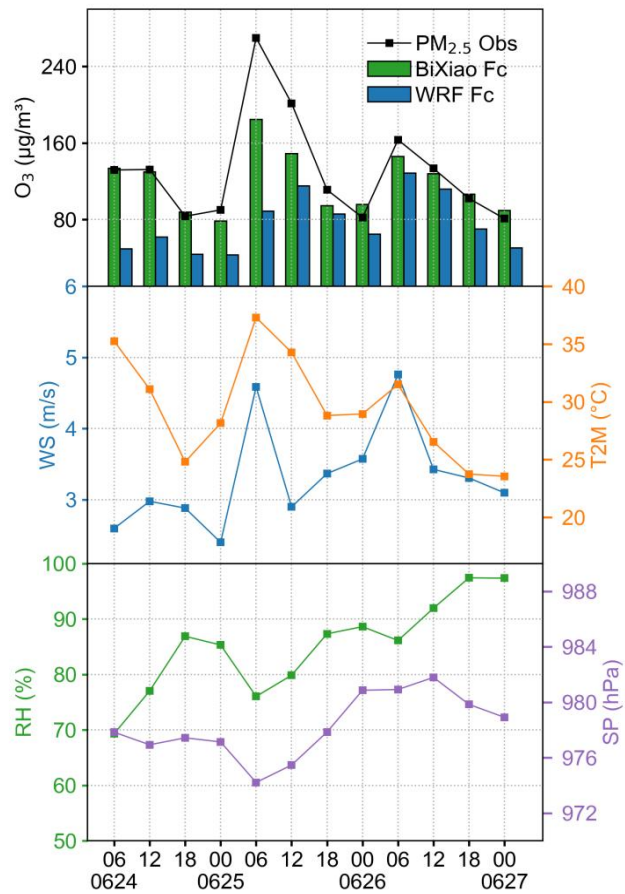

**Figure 8:** Time series of average pollutant concentrations during the $O_3$ pollution event and forecast values from BiXiao and WRF-Chem (top), surface wind speed and temperature time series (middle), and surface relative humidity and pressure time series (bottom).

**Figure 9** shows the average pollution levels across all grids during the entire pollution event, as well as the average forecast mean absolute errors of BiXiao and WRF-Chem in different grids. BiXiao's forecast average mean absolute error is consistently lower than that of WRF-Chem across all grids. Moreover, the distribution of BiXiao's forecast errors is more uniform across different grids, whereas WRF-Chem exhibits larger forecast errors in the central region, where $O_3$ pollution is more severe. Overall, BiXiao outperforms WRF-Chem significantly during this $O_3$ pollution event.

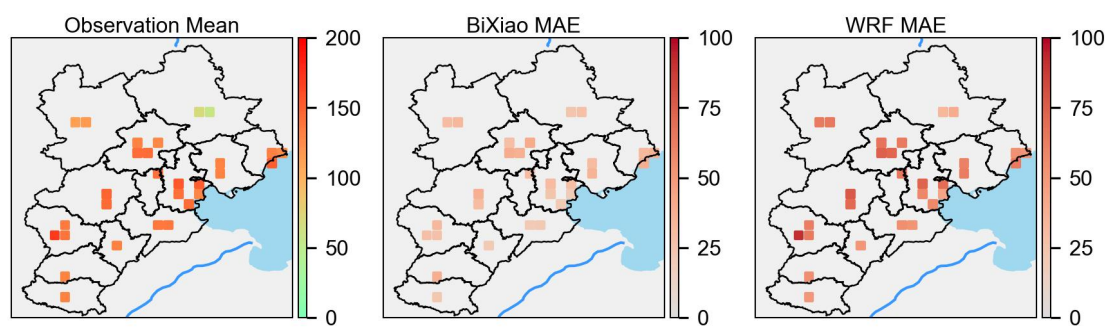

**Figure 9:** Distribution of average $O_3$ observed concentration (left), BiXiao forecast average absolute error (middle), and WRF-Chem forecast average absolute error (right) across different grids for the 6-72 hours forecast period starting from 00:00 UTC, June 24, 2022.

### 5.2.2 PM$_{2.5}$ Pollution Case

From October 29 to November 1, 2023, the BTH region experienced a severe PM$_{2.5}$ pollution event, influenced by both local emissions (such as motor vehicle exhaust, industrial furnaces, and straw burning) and regional transport. Several cities, including Beijing, Tianjin, Shijiazhuang, and Anyang, issued heavy pollution weather warnings. This study selected this pollution episode for a comparative analysis, with a forecast start time of 00:00 UTC, October 29, 2023, and a total forecast period of 72 hours, to evaluate the forecasting capability of the BiXiao model and compare it with WRF-Chem.

Meteorologically, starting from October 29, the BTH region and surrounding areas were under the control of a saddle-shaped pressure pattern, characterized by low surface pressure and persistent southerly winds near the surface with an average wind speed of only 2 m/s. Due to the calm and stable weather pattern, atmospheric diffusion conditions in the region were unfavorable. As shown in **Figure 10**, this pollution event was also accompanied by prolonged high temperatures and humidity. The highest surface temperatures exceeded 23 °C, while the relative humidity at night reached 97%, nearly saturated, which favored the accumulation and transformation of particulate pollution. During this pollution event, the PM$_{2.5}$ concentration peaked at 20:00 local time (12:00 UTC) on both October 29 and 31, with the average grid concentrations in the BTH region reaching 151.4 μg/m³ and 147.1 μg/m³, respectively.

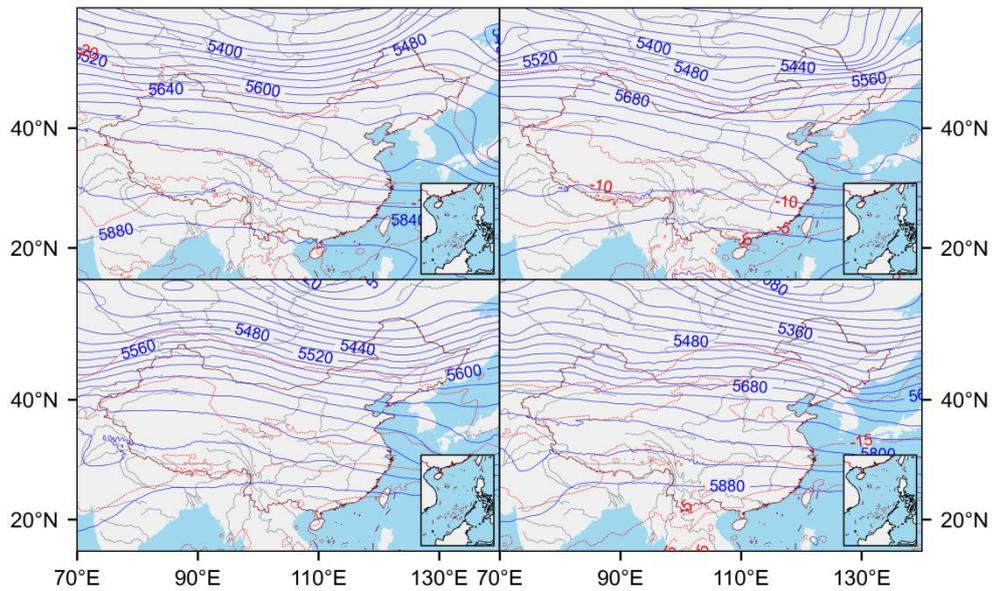

**Figure 10:** 500 hPa geopotential height and temperature map from 00:00 UTC, October 29 to November 1, 2023.

**Figure 11** shows the average observed PM$_{2.5}$ concentrations across all grids during the forecast period, as well as the forecasted concentrations from the BiXiao and WRF-Chem models. Comparing the forecast results every 6 hours with observed concentrations, BiXiao's forecasts are closer to the observed values, with the highest correlation coefficient of 0.82 at the 6h and 30h forecast times. The correlation coefficients are lower at the 24h and 72h forecasts, at 0.49 and 0.56, respectively, with the average absolute error around 37.07±16.65 μg/m³. In contrast, WRF-Chem's forecasts exhibit an underestimation, with lower correlation coefficients, peaking at 0.67 at the 30h forecast time.

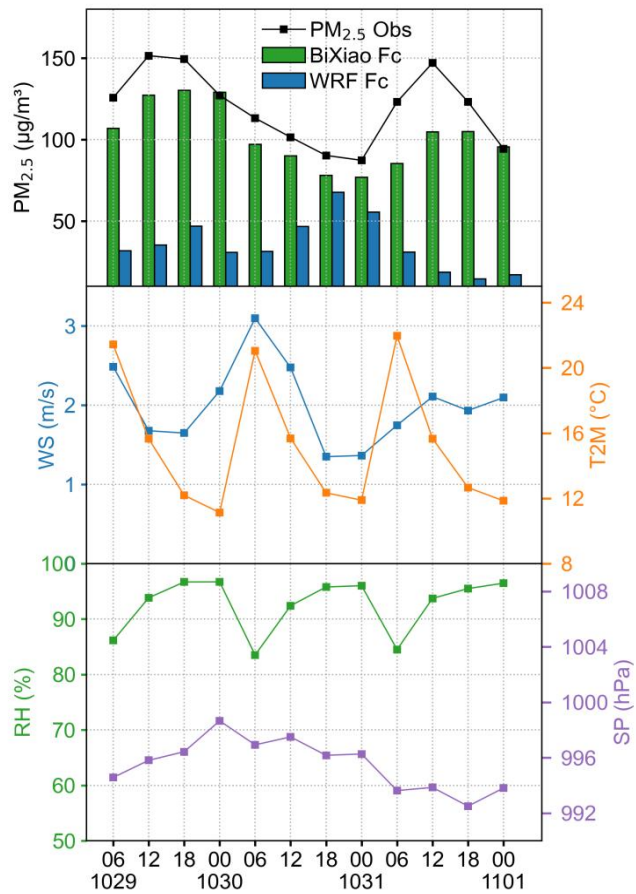

**Figure 11:** Time series of average pollutant concentrations during the PM$_{2.5}$ pollution event and forecast values from BiXiao and WRF-Chem (top), surface wind speed and temperature time series (middle), surface relative humidity and pressure time series (bottom).

**Figure 12**. shows average fine pollutant concentrations in different grids during the forecast period, along with the forecast average mean absolute errors from BiXiao and WRF-Chem in various grids. BiXiao exhibits higher forecast accuracy for fine particulate concentrations in northern regions of the BTH area compared to the southern regions, with the northern average absolute error for PM$_{2.5}$ around 32 μg/m³ (e.g., Zhangjiakou, Beijing, Chengde). Compared to BiXiao, WRF-Chem shows larger forecast errors in most grids, with significantly higher errors in the heavily polluted central and southern regions. Overall, BiXiao performs significantly better than WRF-Chem in forecasting this PM$_{2.5}$ pollution event.

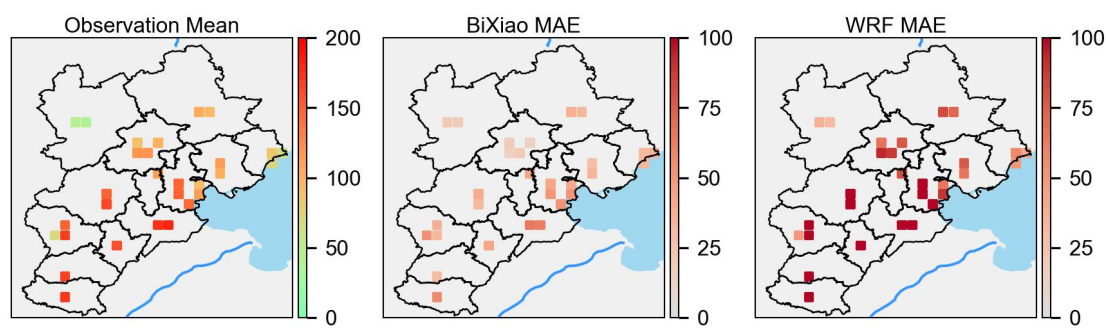

**Figure 12:** Distribution of average PM$_{2.5}$ observed concentrations (left), BiXiao forecast average absolute errors (middle), and WRF-Chem forecast average absolute errors (right) for different grids over the 6-72 hours forecast period, starting from 00:00 UTC on October 29, 2023.

## 6 Conclusion

To address the limitations of traditional numerical models, such as low computational efficiency and insufficient resolution in atmospheric environmental element predictions, this study proposes the "BiXiao" AI model for atmospheric environmental forecasting based on a discontinuous grid design. Leveraging the rapid development of AI-driven meteorological large models, the "BiXiao" model innovatively integrates the weather and environmental modules in a "heterogeneous" architecture. This approach enables the direct use of discrete station observation data, overcoming the current dependency of AI large models on gridded data.

The meteorological module employs a 3D Swin Transformer as the backbone network to extract three-dimensional atmospheric dynamic features. The environmental module combines the evolution characteristics of the meteorological fields with discrete environmental field data to accurately predict six major pollutants. Additionally, the model can perform inference tasks using a single GPU, significantly improving computational efficiency compared to traditional numerical models.

In validation experiments using the test set, the BiXiao model demonstrated strong performance, with the best 6-hour O$_3$ forecast, achieving a correlation coefficient (PCC) of 0.91. The PCC for PM$_{2.5}$ and PM$_{10}$ forecasts were 0.86 and 0.79, respectively. In comparisons with the mainstream numerical model CAMS for 72-hour forecasts, BiXiao outperformed CAMS in the first 48 hours in terms of both PCC and RMSE for particulate matter forecasts. After 48 hours, BiXiao's performance slightly lagged behind CAMS in small regional areas, though it still outperformed CAMS in most regions. The model reduced short-term RMSE errors for PM$_{2.5}$ and PM$_{10}$ by over 50% compared to CAMS.

In case study simulations, the BiXiao model showed better performance in forecasting PM$_{2.5}$ and O$_3$ pollution events, with significantly lower mean average errors compared to the WRF-Chem model, demonstrating stronger robustness. This positions the BiXiao model as a new paradigm for fine-scale urban atmospheric environment forecasting.

The development of the BiXiao model marks an initial attempt to apply AI technologies in atmospheric environmental research. Future extensions will focus on the following directions: Firstly, incorporating satellite remote sensing, mobile monitoring data, and other observational data to build a multi-modal dataset that enhances the model's ability to capture local pollution sources. Secondly, expanding the model's research into pollution source tracing capabilities, in addition to forecasting capabilities. Finally, addressing the simulation needs for different emission reduction scenarios in atmospheric environmental forecasting, incorporating emission source impact factors into the BiXiao model. The model will also be extended from the BTH region to nationwide applications, with a focus on its suitability in complex terrains (such as plateaus and basins) and typical pollution areas (such as the Yangtze River Delta and Pearl River Delta). With these improvements, the BiXiao model is expected to become an important technological tool in the "pollution reduction and carbon reduction" strategy, providing a Chinese solution for global atmospheric environmental governance.